\documentclass[a4paper]{article}

\usepackage[T1]{fontenc}
\usepackage[utf8x]{inputenc}
\usepackage[english]{babel}
\usepackage[font=small,labelfont=bf,labelsep=colon]{caption}
\usepackage[super,sort&compress,comma]{natbib} 
\usepackage{amssymb,amsbsy}
\usepackage[intlimits,sumlimits]{amsmath}

\usepackage{droidsans}
\usepackage[colorlinks=true, allcolors=blue]{hyperref}

\newcommand{\email}[1]{\href{mailto:#1}{\tt{\nolinkurl{#1}}}}
\newcommand{\orcid}[1]{ORCID: \href{https://orcid.org/#1}{\tt{\nolinkurl{#1}}}}

\usepackage[sfdefault,lf]{carlito}

\usepackage{tabulary}
\usepackage{longtable}
\usepackage{makecell}
\usepackage{multirow}
\usepackage{graphicx}
\usepackage{amsfonts,amsmath,amssymb,amsbsy}
\usepackage{dcolumn,bm}
\usepackage{float}
\usepackage{subfig}
\usepackage{url}
\usepackage{color}
\usepackage{gensymb}
\usepackage{hyperref}% add hypertext capabilities
\usepackage[mathlines]{lineno}
\usepackage{makecell}
\usepackage{placeins}
\usepackage[parfill]{parskip}

\usepackage{fancyhdr}
\usepackage{authblk}
\usepackage[a4paper,top=3cm,bottom=2cm,left=3cm,right=3cm,marginparwidth=1.75cm]{geometry}

\usepackage{amsmath}
\usepackage{graphicx}
\usepackage{booktabs}
\usepackage{color,soul}

\usepackage[figuresright]{rotating}

\title{Chelation of the mercury ions by polyethyleneimine: Atomistic molecular dynamics study}
\date{}

\author[1,2]{Halyna Butovych}
\author[1,3]{Jaroslav Ilnytskyi}
\author[2]{Erkki L\"{a}hderanta}
\author[1,3]{Taras Patsahan}

\affil[1]{Institute for Condensed Matter Physics of the National Academy of Sciences of Ukraine, 1 Svientsitskii str., 79011, Lviv, Ukraine}
\affil[2]{Department of Physics, School of Engineering Science, LUT University, Yliopistonkatu 34, FI-53850 Lappeenranta, Finland}
\affil[3]{Institute of Applied Mathematics and Fundamental Sciences, Lviv Polytechnic National University, 12 S. Bandera Str., 79013 Lviv, Ukraine}
\affil[*]{Corresponding author: \email{tarpa@icmp.lviv.ua}}

\begin{document}
\maketitle

\begin{abstract}
\noindent
Contamination of water by heavy metal ions represents a significant environmental concern. Among various remediation methods, chelation has proven to be an effective technique in water treatment processes. This study investigates the chelating properties of linear polyethyleneimine (PEI) and its complexation with divalent mercury ions (Hg$^{2+}$) in aqueous solution. Atomistic molecular dynamics (MD) simulations were carried out using the OPLS/AA force field to examine the microscopic structure of PEI–Hg$^{2+}$ complexes. PEI chains of varying lengths were considered, and it was found that a single linear PEI molecule containing ten amino groups is capable of coordinating up to four Hg$^{2+}$ ions. The stability of the resulting complexes was further supported by density functional theory (DFT) calculations.
\end{abstract}

%\begin{keyword}
%%% keywords here, in the form: keyword \sep keyword
%%% MSC codes here, in the form: \MSC code \sep code
%%% or \MSC[2008] code \sep code (2000 is the default)
%chelation \sep polyethylenimine  \sep mercury \sep molecular dynamics 
%\end{keyword}

\newcommand{\vvec}[1]{\mathbf{#1}}

%%%%%%%%%%%%%%%%%%%%%%%%%%%%%%%%%%%
\section{Introduction}\label{sec:1}
%%%%%%%%%%%%%%%%%%%%%%%%%%%%%%%%%%%

Contamination of the natural environment with toxic heavy metal ions, particularly mercury (Hg$^{2+}$), remains a critical concern~\cite{WagnerDobler2000, Driscoll2013, Gworek2020}, yet their efficient removal poses a significant challenge. Key priorities in the development of modern purification technologies~\cite{Singh2018, Qasem2021} include enhancing removal efficiency, minimizing associated costs, and avoiding the formation of toxic by-products. Among the most effective strategies is adsorption~\cite{Singh2018}, which frequently relies on chelation, where a specific chemical moiety (a chelate) forms a stable, ring-like complex with one or more heavy metal ions. The practical implementation of this technique in water treatment systems often involves magnetic nanoparticles functionalized with chelator-containing polymers~\cite{Farrukh2013, Moja2021}. The resulting complex, comprising the nanoparticle and the polymer-encapsulated metal ions, can be removed from the solution via an external magnetic field. The efficiency of this process can be further enhanced by optimizing the polymer architecture and selecting the most effective chelating agent. The latter depends critically on a detailed understanding of the complexation process between the chelator and metal ions, which is the focus of the present study.
Various chemical compounds have been employed for the chelation of Hg$^{2+}$ ions, including ethylenediaminetetraacetic acid (EDTA)~\cite{BlaurockBusch2020}, and 2,3-dimercaptosuccinic acid (DMSA)\cite{Singh2018}. Polyethylenimine (PEI) satisfies these criteria well. It is biocompatible and can be readily synthesised in both linear and branched forms~\cite{Kobayashi1987, Jia2014, Zhang2019}, including hyperbranched architectures~\cite{Kuo2001, Nam2015}. PEI has been widely used to modify inorganic materials and both synthetic and natural polymers, exhibiting an excellent regeneration capacity~\cite{Ayalew2022}. Representative examples include PEI-modified graphene oxide hydrogels~\cite{Arshad2019}, magnetic nanoparticles~\cite{Shen2013}, chitosan/PEI magnetic hydrogels~\cite{Chen2022}, and PEI-functionalised magnetic cellulose nanocrystals~\cite{Zhu2022}. These characteristics underscore the versatility and efficiency of PEI as a component in chelation-based water purification systems.

The coordination of PEI with heavy metal ions in aqueous solutions is rather complex and remains insufficiently understood. This calls for broader application of theoretical approaches and computer simulations, although existing studies have primarily focused on investigating the properties of aqueous solutions of PEI molecules without considering the presence of heavy metal ions.
Therefore, atomistic and coarse-grained molecular dynamics simulations have been applied to investigate the structural and protonation-dependent properties of polyethyleneimine (PEI). Ziebarth and Wang~\cite{ziebarthUnderstandingProtonationBehavior2010} combined all-atom simulations with Monte Carlo titration using a coarse-grained model that incorporates screened Coulomb interactions to describe the protonation behaviour of linear PEI. Gallops et al.~\cite{gallopsEffectProtonationLevel2019a} employed atomistic simulations 
of a 40-mer linear PEI to explore the effects of varying protonation levels and salt concentrations on chain conformation. 
Choudhury and Roy~\cite{choudhuryStructuralDynamicalProperties2013a} also used all-atom molecular dynamics to examine PEI in explicit water and different protonation states, with particular focus on solvation shell dynamics. Beu and co-workers developed both CHARMM-based atomistic force fields and MARTINI coarse-grained models for linear and branched PEI, enabling efficient simulations of PEI–DNA condensation and the size- and protonation-dependent behaviour of PEI in solution~\cite{beuCHARMMForceField2018,beuAtomisticCoarsegrainedModeling2019,Beu2019b,BeuBranchedPolyethyleneimineCHARMM2022}.
Mintis et al.~\cite{mintisEffectPHMolecular2020} conducted detailed atomistic molecular dynamics simulations to investigate how pH and molecular weight influence the structure, hydration, and dynamics of linear and branched PEI, revealing distinct conformational preferences, stiffness, and diffusion behaviours as a function of ionization degree and chain architecture.
Besides classical approaches, the recent quantum computational calculations based on the densitity functional theory (DFT) have provided valuable insights into the structural and electronic properties of polyethyleneimine~(PEI). 
Buijs~\cite{buijsCOCapturePEI2023} investigated the oxidation stability of linear and branched PEI, identifying the formation of 
$\alpha$-amino hydroperoxides as the rate-determining step under ideal conditions relevant to CO$_2$ capture applications. 
In another study, Hussan et al.~\cite{Hussan_2025} combined DFT with spectroscopic techniques to explore charge transport mechanisms, hydrogen bonding networks, and water solubility in PEI. 

Among the force fields available for describing the interactions of PEI with ions and various organic and inorganic compounds, OPLS/AA~\cite{jorgensenDevelopmentTestingOPLS1996} has proven to be particularly effective~\cite{kwolek2016interactions, li2016understanding, tanis2021molecular}. It has also been employed in studies on the removal of Hg$^{2+}$ ions from water using stacked graphene membranes, graphene oxide, and functionalized single-walled carbon nanotubes~\cite{giri2021heavy, han2021enhanced, anitha2015removal}, with interaction parameters as proposed in~\cite{babu2006empirical}.

The aim of this study is to advance our understanding of how exactly linear PEI molecules form complexes with mercury ions, 
as this is a key factor in improving existing PEI-based wastewater purification systems. 
To this end, we apply a PEI model based on the OPLS/AA force field and perform molecular dynamics (MD) simulations to examine the structure and stability of PEI-Hg$^{2+}$ complexes in water. We focus on the structure of these complexes for a range of relatively short PEI chains and varying numbers of Hg$^{2+}$ ions. The stability of the complexes obtained from the MD simulations is further assessed using density functional theory (DFT) calculations. To the best of our knowledge, there are currently no studies 
on PEI-Hg$^{2+}$ systems employing the OPLS/AA force field.

The outline of the study is as follows. Section~\ref{sec:2} presents the all-atom model of PEI, and the MD simulation protocol is described in Section~\ref{sec:3}. Section~\ref{sec:4} reports the results of the MD simulations in terms of the structural characteristics of the complexes formed by a single linear PEI chain of various lengths and Hg$^{2+}$ ions. 
In Section~\ref{sec:5}, these results are validated using density functional theory (DFT) calculations. The study concludes with a summary of the main findings.

%%%%%%%%%%%%%%%%%%%%%%%%%%%%%%%%%%%%%%%%%%%%%%%%%%%%%%%%%%%%%%%%%%%%
\section{All-atom model of linear PEI chains based on the OPLS/AA force field} \label{sec:2}
%%%%%%%%%%%%%%%%%%%%%%%%%%%%%%%%%%%%%%%%%%%%%%%%%%%%%%%%%%%%%%%%%%%%

\begin{figure}[htb]
	\begin{center}
		\includegraphics[width=0.7\linewidth]{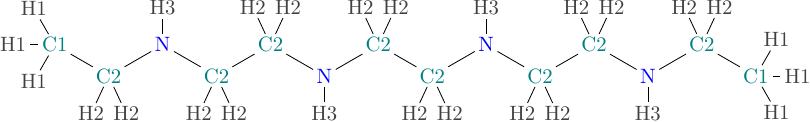}
		\caption{\label{fig:pei_oplsaa} Atomistic model of PEI molecule containing $4$ nitrogen atoms (referred to as PEI-4) described using the OPLS/AA force field (see parameters in Table~\ref{tab:pei_oplsaa}).}
	\end{center}
\end{figure}

Molecular dynamics (MD) simulations have been performed using atomistic models of the systems containing a complex formed by a single PEI molecule and ions of mercury (Hg$^{2+}$) in an aqueous environment. The systems were studied in the isothermal-isobaric (NPT) ensemble at the temperature of $298.15$~K and pressure of $1$~atm. 
The \textit{LAMMPS} software package \cite{Thompson_2022} of version \textit{2Aug2023} is applied to run MD simulations, while the \textit{moltemplate} software \cite{Jewett_2021} is used to build a molecule of PEI and to generate input files with its structure and topology. 
For the model of PEI molecule (Fig~\ref{fig:pei_oplsaa}), the conventional OPLS/AA force field \cite{jorgensenDevelopmentTestingOPLS1996} is used,
with the harmonic potential $E = K(r-r_0)^2$ for directly bonded atoms ($r_0$ is equilibrium bond distance), the harmonic angle potential $E = K (\theta - \theta_0)^2$ ($\theta_0$ is the equilibrium value of the angle), and the dihedral potential $E = \frac{1}{2} K_1 [1 + \cos(\phi)] + \frac{1}{2} K_2 [1 - \cos(2 \phi)] + \frac{1}{2} K_3 [1 + \cos(3 \phi)] + \frac{1}{2} K_4 [1 - \cos(4 \phi)]$ ($\phi$ is dihedral angle) to describe intra-molecular interactions. 
For non-bonded interactions we use the Lennard-Jones 12-6 potential (LJ), $E = 4 \epsilon \left[ \left(\frac{\sigma}{r}\right)^{12} - \left(\frac{\sigma}{r}\right)^6 \right]$ and Coulombic pairwise interaction $E = \frac{q_i q_j}{ r}$. All parameters for bonded and non-bonded interactions, charges and masses of PEI atoms are listed in Table~\ref{tab:pei_oplsaa}. In Fig.~\ref{fig:pei_oplsaa} one can see an example representing the chemical structure of PEI molecule consisting of $4$~nitrogen atoms. 
In our study we consider the PEI molecules of different lengths containing $4$, $5$ and $10$ nitrogen atoms, which are denoted by PEI-4, PEI-5 and PEI-10, respectively. The water molecules are described using the rigid SPC/E model~\cite{Berendsen_1987}. For the mercury ion Hg$^{2+}$ the force field parameters are applied as suggested in \cite{giri2021heavy,han2021enhanced,anitha2015removal}, which are considered to be consistent with the OPLS/AA and SPC/E models. 

\begin{table}
\begin{center}
\caption{\label{tab:pei_oplsaa} OPLS/AA force field parameters for PEI molecule and Hg$^{2+}$ ion. Notations for the PEI atoms correspond to those used in Fig.~\ref{fig:pei_oplsaa}.}
\begin{tabular}{lll} 
\begin{tabular}[t]{lllll}
	\hline
	{Type} & {Charge} & Mass   & $\epsilon$,$\frac{\text{kcal}}{\text{mol}}$      & $\sigma$,\AA{}   \\ \hline
    N    & -0.78    & 14.007& 0.17  & 3.3 \\ %731
	C1  & -0.18    & 12.011 & 0.066 & 3.5\\ %80
	C2  & 0.08     & 12.011 & 0.066 & 3.5\\ %737
	H1  & 0.06     & 1.008  & 0.03  & 2.5 \\ %85
	H2  & 0.06     & 1.008  & 0.015 & 2.5 \\ %741
	H3  & 0.38     & 1.008  & 0     & 0 \\ %740
	Hg$^{2+}$  &  2    & 200.592  & 0.0409  & 2.36 \\ %740
	\hline
\end{tabular}
&\quad &
\begin{tabular}[t]{lll}
	\hline
	{Bond}  & K, $\frac{\text{kcal}}{\text{mol} \cdot \text{A}^2}$ & $r_0$, \AA{} \\ \hline
	C-C & 268                       & 1.529    \\ %737-737,80-737
	C-N & 382                       & 1.448    \\ %731-737
	C-H & 340                       & 1.09     \\ %737-741,80-85
	N-H & 434                       & 1.01     \\ %731-740
	\hline
\end{tabular}
\end{tabular}
\quad\\
\vspace{0.2cm} 
\begin{tabular}{lll}
\begin{tabular}[t]{lll}
	\hline
	{Angle}     & K, $\frac{\text{kcal}}{\text{mol} \cdot \text{rad}^2}$ & ${\theta_0}\degree$ \\ \hline
	C-C-N & 56.2                      & 109.47               \\ %737-737-731
	H-C-H & 33.0                      & 107.8                \\ %741-737-741
	N-C-H & 35.0                      & 109.5                \\ %731-737-741
	C-C-H & 37.5                      & 110.7                \\ %737-737-741
	C-N-C & 51.8                      & 107.2                \\ %737-731-737
	C-N-H & 35.0                      & 109.5                \\ %737-731-740
	\hline
\end{tabular}
&\quad&
\begin{tabular}[t]{lllll}
	\hline
	{Dihedral}      &   ${\text{K}}_1$,$\frac{\text{kcal}}{\text{mol}}$     &   ${\text{K}}_2$,$\frac{\text{kcal}}{\text{mol}}$     &  ${\text{K}}_3$,$\frac{\text{kcal}}{\text{mol}}$     &  ${\text{K}}_4$,$\frac{\text{kcal}}{\text{mol}}$ \\ \hline
	N-C-C-N & 11.035 & -0.968 & 0.27  & 0 \\ %731-737-737-731
	N-C-C-H & -1.013 & -0.709 & 0.473 & 0 \\ %731-737-737-741
	H-C-C-H & 0      & 0      & 0.3   & 0 \\ %741-737-737-741
	C-C-N-C & 0.416  & -0.128 & 0.695 & 0 \\ %737-737-731-737
	C-C-N-H & -0.19  & -0.417 & 0.418 & 0 \\ %737-737-731-740
	H-C-N-C & 0      & 0      & 0.56  & 0 \\ %741-737-731-737
	H-C-N-H & 0      & 0      & 0.4   & 0 \\ %741-737-731-740
	\hline
\end{tabular}
\end{tabular}
\end{center}
\end{table}

%%%%%%%%%%%%%%%%%%%%%%%%%%%%%%%%%%%%%%%%%%%%%%%%%%%%%%%%%%%%%%%%%%%%
\section{Computer simulation details} \label{sec:3}
%%%%%%%%%%%%%%%%%%%%%%%%%%%%%%%%%%%%%%%%%%%%%%%%%%%%%%%%%%%%%%%%%%%%

In all MD simulations performed in this study, periodic boundary conditions were applied in three spatial dimensions. The particle–particle particle–mesh (PPPM) scheme was employed to compute long-range electrostatic interactions with an accuracy of $10^{-4}$. A cutoff radius of $12$~\AA{} was used for Coulombic interactions, while the Lennard-Jones interactions were truncated at a distance of $10$~\AA{}.
The SHAKE algorithm is applied to constrain the bonds and angle in the rigid model of SPC/E water molecule \cite{Berendsen_1987}. The geometrical combination rule is used for the Lennard-Jones cross-interactions of all atoms in the system. Intramolecular non-bonding interactions between only $1-4$ neighbouring atoms are considered, and as suggested in the OPLS/AA model, they are scaled by a factor of $0.5$.

\begin{figure}[!tbh]
	\centering
	\includegraphics[width=0.4\linewidth]{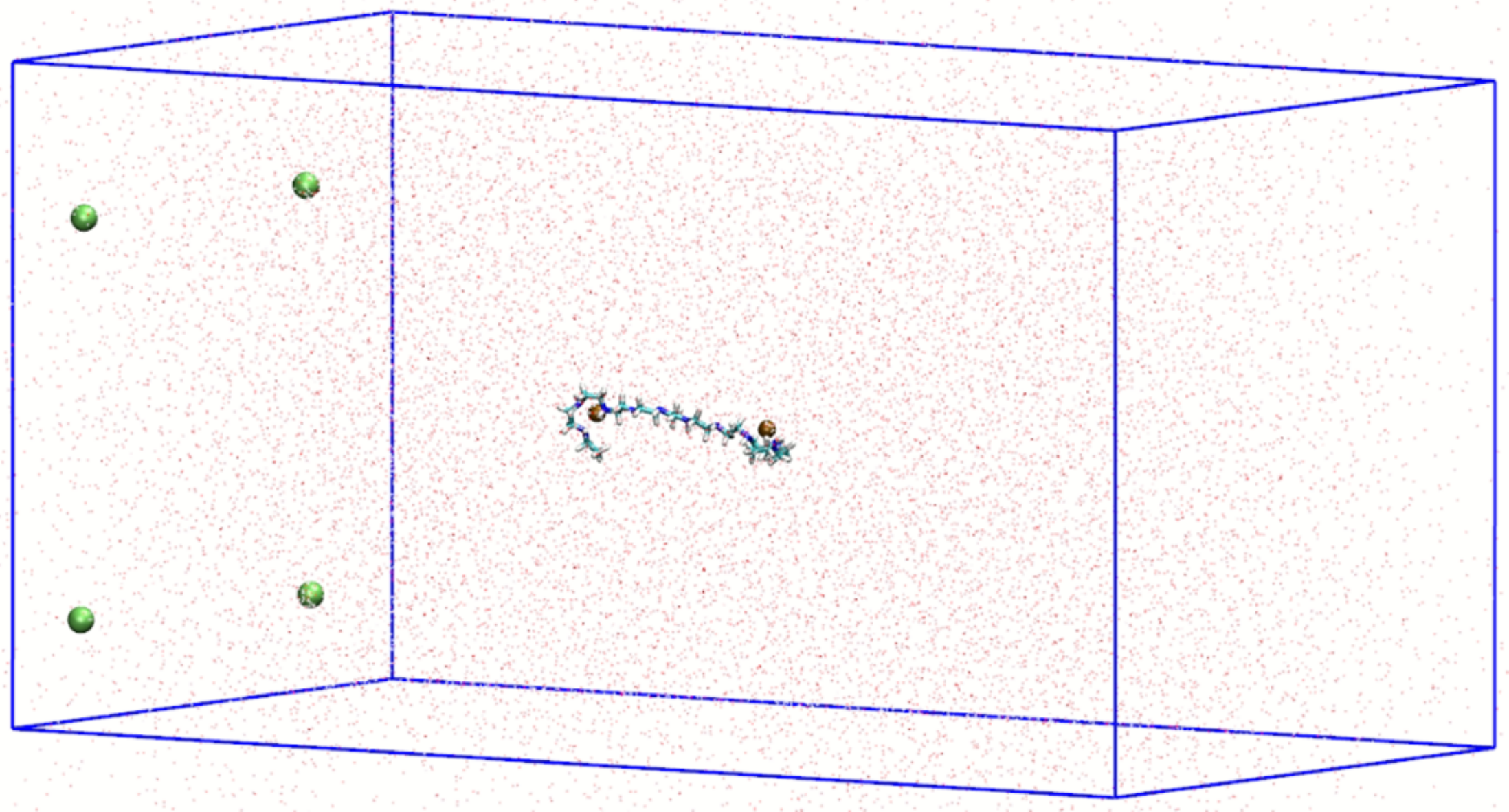} \qquad \qquad
	\caption{Simulation box with a PEI-10 + 2~Hg$^{2+}$ ions complex (in the middle) and 4~counterions (at the boundary). Water molecules are seen as small red dots. 
 }
	\label{fig:boxPEI}
\end{figure}

The simulations comprised four sequential stages: 
1)~formation of a PEI-Hg$^{2+}$ complex in a vacuum, i.e. without water molecules; 2) then, the complex of PEI-Hg$^{2+}$ is introduced in the water environment and the volume of the simulation box is relaxed at the ambient conditions; 3)~equilibration of the system obtained at the previous stage; 4) and the production run is performed at the last stage. Stage~1 and Stage~2 are the preparation stages needed to initialize a system of our interest in the most rapid and efficient way. Below, we describe each of the stages in detail.

\begin{figure}[!htb]
	\centering
\subfloat[PEI-4 + 1~Hg$^{2+}$]{\label{fig:PEI_model4}{\includegraphics[width=0.18\linewidth]{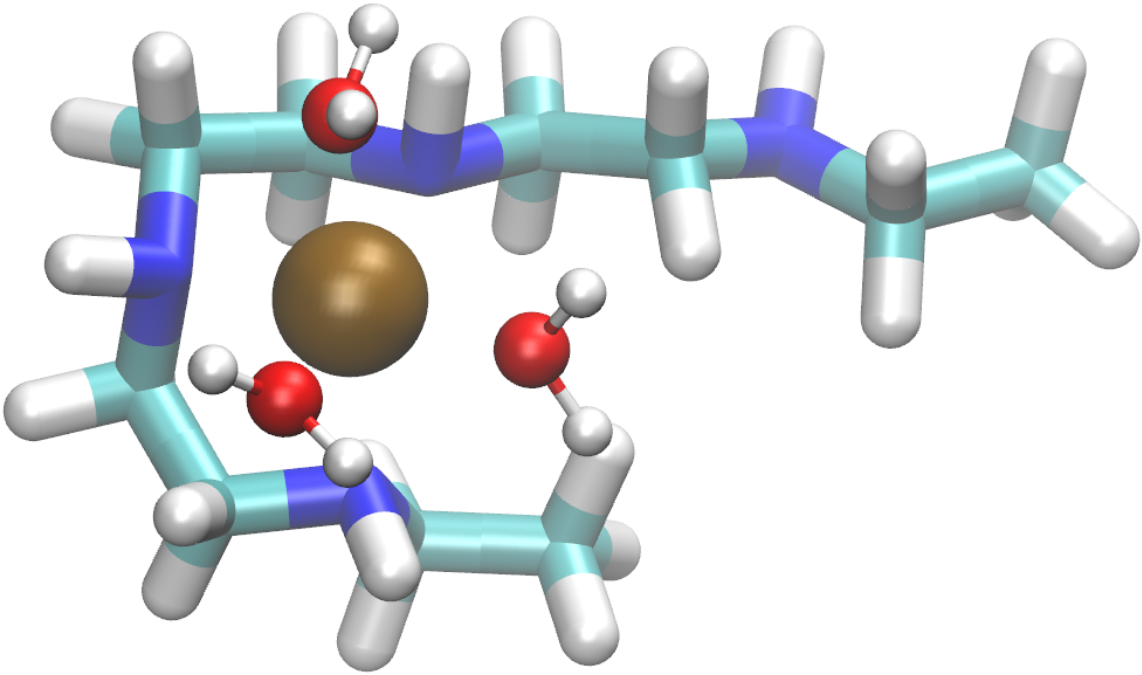} }}
\subfloat[PEI-5 +1~Hg$^{2+}$]{\label{fig:PEI_model5}{\hspace{1em}\includegraphics[width=0.112\linewidth]{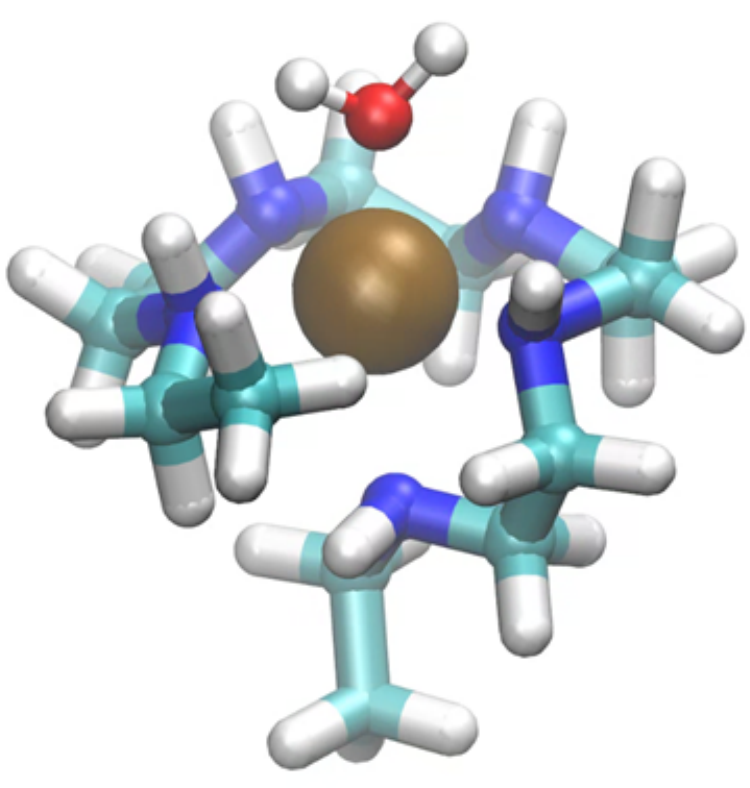} }\hspace{1em}}
\subfloat[PEI-10 + 1~Hg$^{2+}$]{\label{fig:PEI_model10_1}{
		\includegraphics[width=0.41\linewidth]{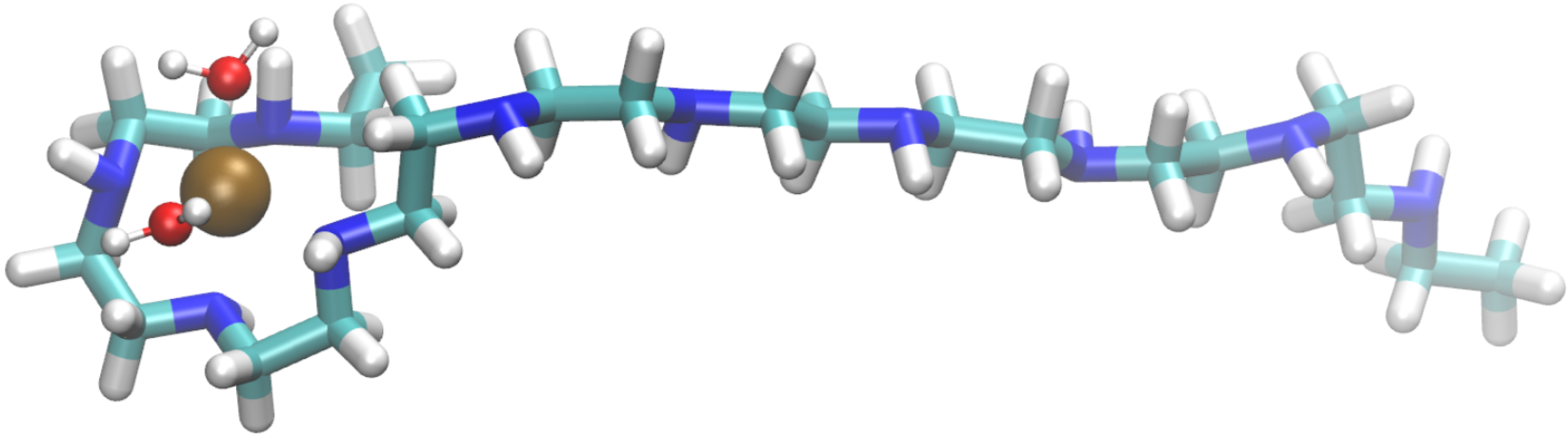} }}
\\ 
\subfloat[PEI-10 + 2~Hg$^{2+}$]{\label{fig:PEI_model10_2}{
		\includegraphics[width=0.425\linewidth]{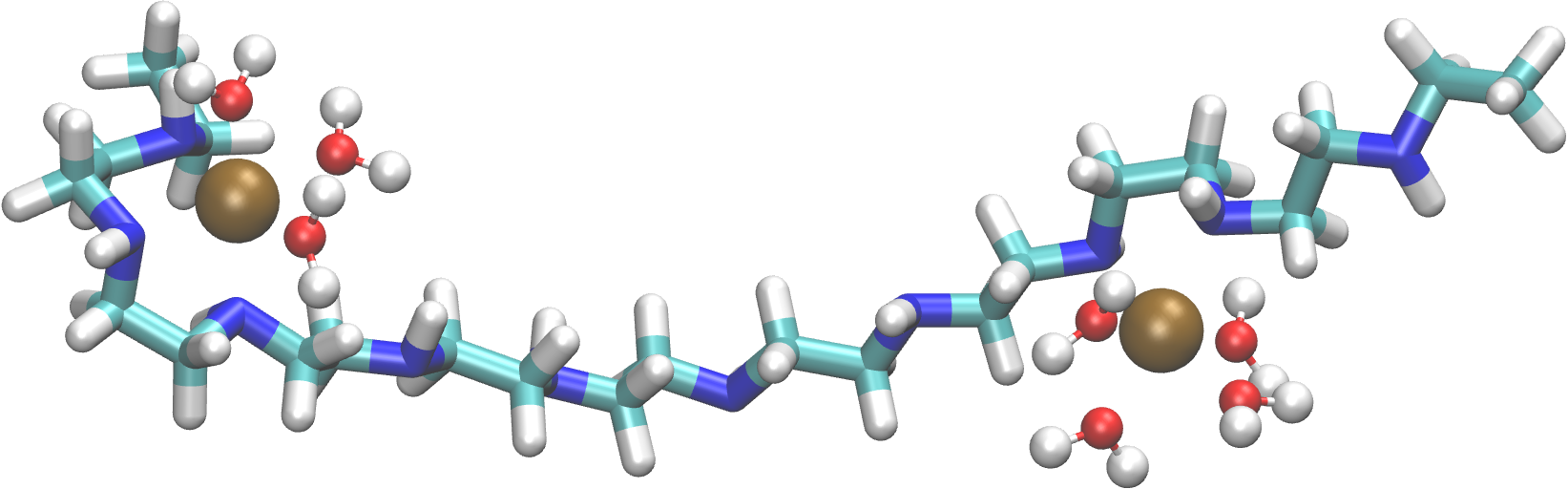} }} \\
\subfloat[PEI-10 + 3~Hg$^{2+}$]{\label{fig:PEI_model10_3}{\includegraphics[width=0.444\linewidth]{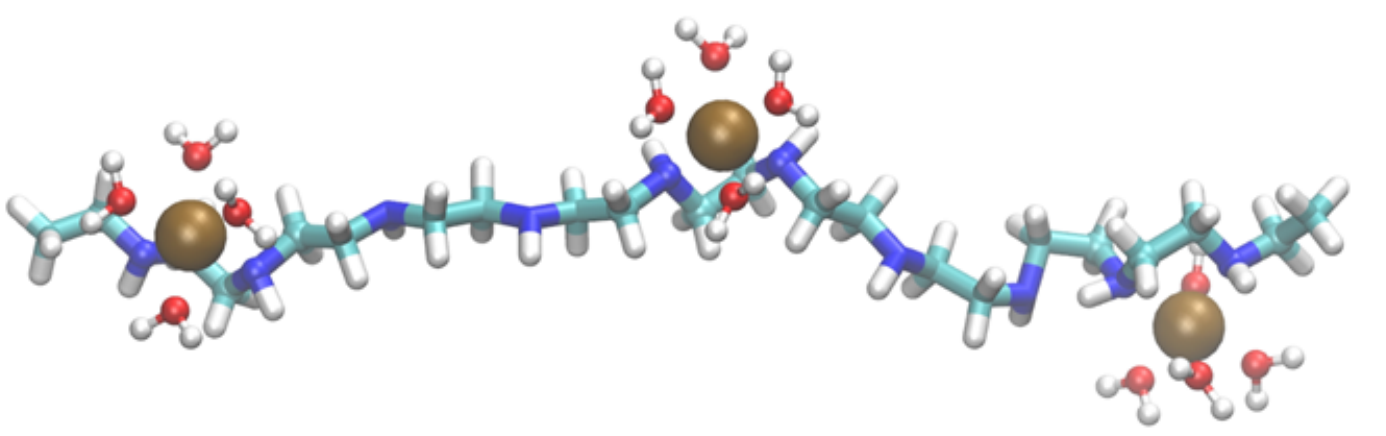} }} 
\subfloat[PEI-10 + 4~Hg$^{2+}$]{\label{fig:PEI_model10_4}{\includegraphics[width=0.389\linewidth]{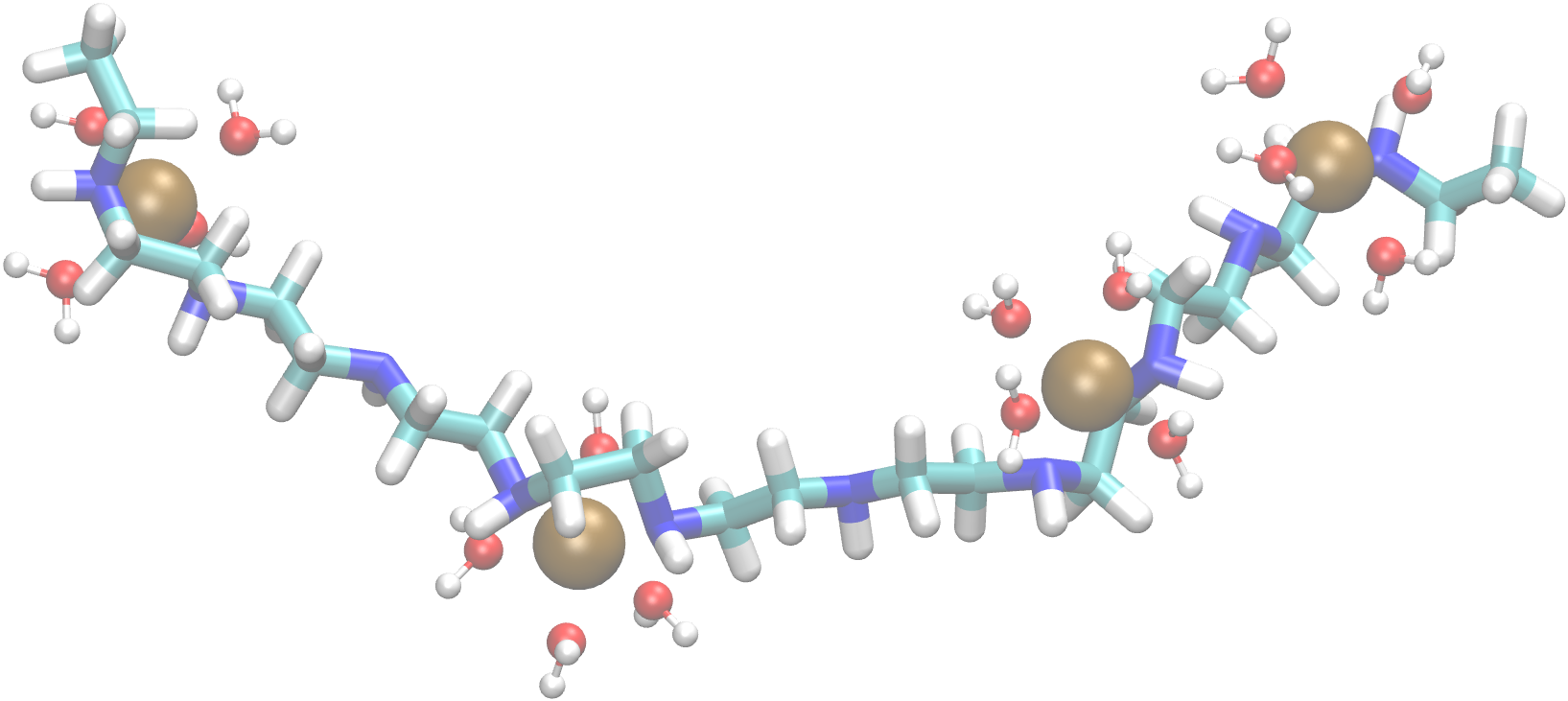} }}  
	\caption{Examples of the PEI + Hg$^{2+}$ complex with water molecules in the first hydration shell. Carbon atoms are shown in cyan, nitrogen atoms -- in blue, hydrogen atoms -- in light-grey, mercury ion -- in brown, oxygen and hydrogen atoms of water molecules are shown as red and light-grey spheres respectively. \label{fig:PEI_models}}
\end{figure}

At the first stage (Stage~1), the PEI molecule and the Hg$^{2+}$ ion are placed in the simulation box in close proximity to each other in such a way as to potentially form a stable complex. There are no water molecules at this stage. The simulation box is of the rectangular shape of size 
$L_{x}=150$~\AA{}, $L_{y}=L_{z}=75$~\AA{}. To preserve the condition of electroneutrality in the system, two counterions were placed in the simulation box at a distance to Hg$^{2+}$ not less than $70$~\AA{} (Fig.~\ref{fig:boxPEI}). The MD simulation was run for this system in the constant volume (NVT ensemble) and at the temperature $298.15$~K to get a complex in a vacuum. At this stage, all ions are fixed (frozen) at their initial positions to avoid approaching 
each other during this simulation run, what can happen in the absence of the screening effect in vacuum, where the Coulomb interaction between ions can be significant 
even at large distances.
We have noticed that a stable complex easily forms within $1$~ns of simulation time~(Fig.~\ref{fig:boxPEI}). The configuration obtained at Stage~1 is used 
in the next stage (Stage~2), at which $20000$ water molecules were inserted randomly. The insertion procedure prevented overlaps between atoms at the minimum distance of $1.5$~\AA{}. Then, the NPT (the pressure is fixed and the volume can vary) simulation was run  at the temperature $298.15$~K and pressure $P=1$~atm. 
We have noticed that $100$~ps was an appropriate period for the simulation box to attain the box size inherent to the given conditions. At this stage, the Hg$^{2+}$ ion could move freely, while the counterions remained fixed at their positions, although they could be rescaled together with changing the size of the simulation box. 
The final size of the box was obtained as $L_{x}=134$~\AA{}, $L_{y}=L_{z}=67$~\AA{}~(see~Fig.~\ref{fig:boxPEI}), which appeared to be still large enough to keep the Hg$^{2+}$ and counterions at a sufficient distance from each other (more than $50$~\AA{}), thus their interactions are considered negligible and the presence of counter ions should not affect 
the complex of PEI and Hg$^{2+}$ ions.
The water density observed at the end of this stage agrees with the well-known value for SPC/E water in the bulk ($\rho=0.99$~g/cm$^{3}$).

\begin{figure}[!htb]
	\centering
	\includegraphics[width=0.4\linewidth]{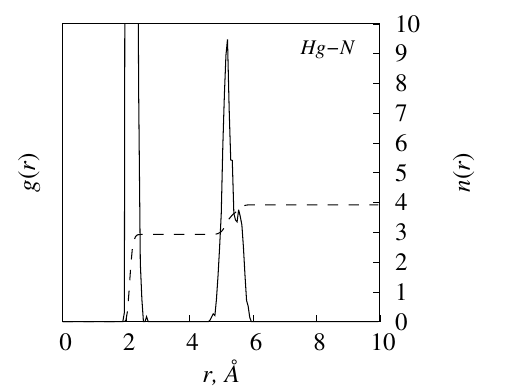}
 	\includegraphics[width=0.4\linewidth]{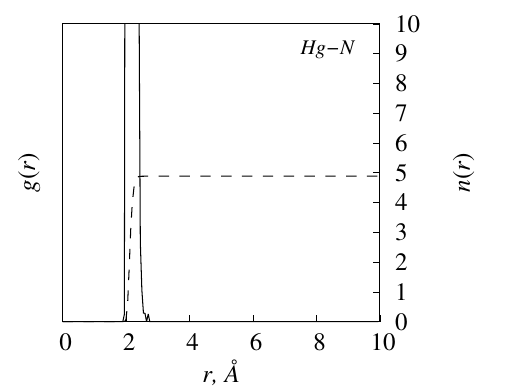}
	\caption{Radial distribution functions $g(r)$ (solid line) and running coordination numbers $n(r)$ (dashed line) Hg-N for PEI-4 + 1~Hg$^{2+}$ (left) and PEI-5 + 1~Hg$^{2+}$ (right) complexes.}
	\label{fig:rdf_PEI4_Hg-N}
\end{figure}

\begin{figure}[!htb]
	\centering
	\includegraphics[width=0.4\linewidth]{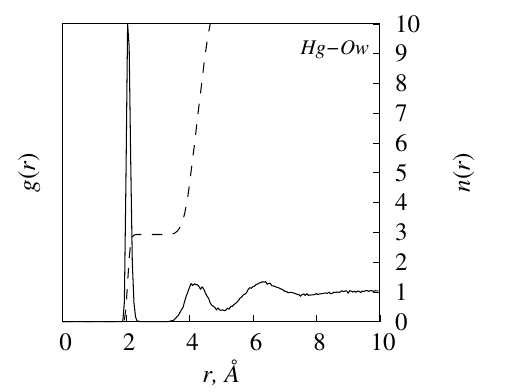}
 	\includegraphics[width=0.4\linewidth]{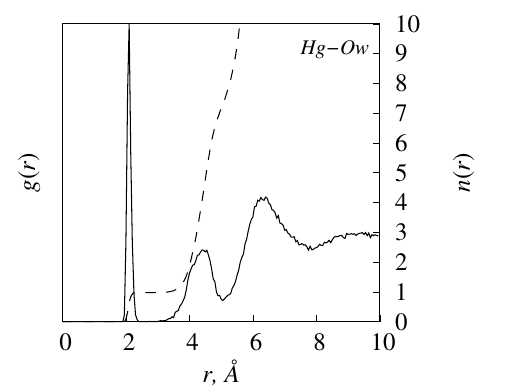}
	\caption{Radial distribution functions $g(r)$ (solid line) and running coordination numbers $n(r)$ (dashed line) Hg-Ow for PEI-4 + 1~Hg$^{2+}$ (left) and PEI-5 + 1~Hg$^{2+}$ (right) complexes.}
	\label{fig:rdf_PEI4_Hg-Ow}
\end{figure}

For the MD simulations performed in constant volume (NVT) at Stage~1, the temperature relaxation parameter was set to $0.01$~ps for the Nos\'{e}-Hoover thermostat,  
while the timestep was $0.5$~fs. 
At Stage~2, the simulations ran at constant pressure (NPT) with the timestep of $1.0$~fs and the Nos\'{e}-Hoover thermostat-barostat was applied with the relaxation parameters $0.1$~ps and $1$~ps for the temperature and pressure, respectively. The simulations at Stage~3 and 4 were performed with the same parameters as in Stage~2.

Since in the result of Stage~1 and Stage~2 the PEI-Hg$^{2+}$ complex is formed and located in water of the proper density at the ambient conditions, the equilibration process can be started. In all cases the equilibration period was $3$~ns. After the system was equilibrated, 
the production run was conducted during $4$~ns and the trajectory was stored at each $1$~ps. It is worth noting, that during the equilibration and the production run
we have been checking the positions of PEI polymer and Hg$^{2+}$ ions to ensure they are still in a complex and whether they did not approach the couterions.
We noticed that the complex was moving randomly with no preferential direction. Therefore, one can conclude that there was no significant interaction between
the Hg$^{2+}$ ions and the counterions, even when the distance between them was getting shorter for some period.
Obviously, there was still a probability that the complex could finally encounter one of the counterions after some time, but it required more time for this to take place.
The simulation trajectories stored during the production run were used to analyze the structure of PEI+Hg$^{2+}$ complexes, in particular, to calculate the radial distribution functions 
between atoms of PEI polymer, Hg$^{2+}$ ions and water molecules.

\section{Molecular dynamics results}\label{sec:4}

According to the protocol described above, we performed a series of MD simulations for PEI-Hg$^{2+}$ complexes in water.
We considered different lengths of PEI chains containing $4$ (PEI-4), $5$ (PEI-5) and $10$ (PEI-10) nitrogen atoms (Fig.~\ref{fig:PEI_models}).
Also we have examined different numbers of Hg$^{2+}$ ions interacting with the longer polymer chain of PEI-10.
Overall, the following systems are studied: 1) PEI-4 + Hg$^{2+}$; 2) PEI-5 + Hg$^{2+}$; 3) PEI-10 + Hg$^{2+}$; 4) PEI-10 + 2 Hg$^{2+}$; 
5) PEI-10 + 3 Hg$^{2+}$; 6) PEI-10 + 4 Hg$^{2+}$; and 7) PEI-10 +  5 Hg$^{2+}$. 
We have verified the stability of these complexes in an aqueous environment. The structural properties in the form of radial distribution functions~$g(r)$ and the running coordination numbers~$n(r)$ are calculated during Stage~4 of the production run. 
The radial distribution functions show the preferred distances between atoms relative to each other. By analysing the running coordination numbers, we can estimate the number of atoms in the first coordination shell of the Hg$^{2+}$ ion.
We observed that the listed complexes exhibited substantial stability during the whole simulation period, except for the complex of PEI-10 with 5~Hg$^{2+}$ ions. 
It indicates that in the case of $5$ or more Hg$^{2+}$ ions arranged on a single PEI-10 polymer, the short distances between these ions lead 
to significant Coulombic repulsion, thereby preventing the formation of a stable complex. Henceforth, we exclude from consideration the complexes involving 
the PEI-10 polymer with more than $4$ Hg$^{2+}$ ions.

\begin{figure}[!htb]
	\centering
	\includegraphics[width=0.4\linewidth]{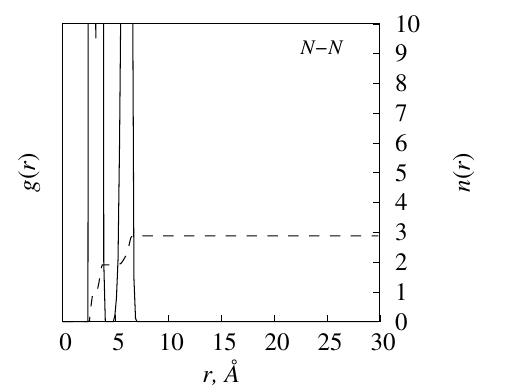}
	\includegraphics[width=0.4\linewidth]{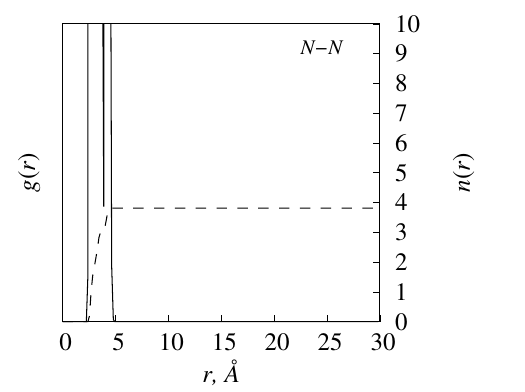}
	\caption{Radial distribution functions $g(r)$ (solid line) and running coordination numbers $n(r)$ (dashed line)  N-N for PEI-4 + 1~Hg$^{2+}$ (left) and PEI-5 + 1~Hg$^{2+}$ (right) complexes.}
	\label{fig:rdf_PEI4_N-N}
\end{figure}

\begin{figure}[!htb]
	\centering
	\includegraphics[width=0.4\linewidth]{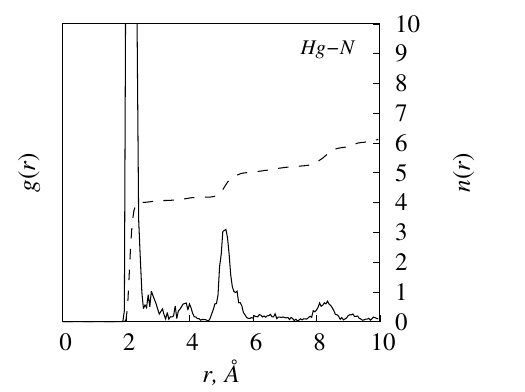}
 	\includegraphics[width=0.4\linewidth]{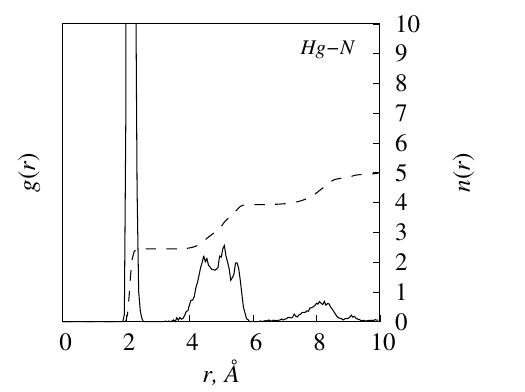}
	\caption{Radial distribution functions $g(r)$ (solid line) and running coordination numbers $n(r)$ (dashed line) Hg-N for PEI-10 + 1~Hg$^{2+}$ (left) and PEI-10 + 2~Hg$^{2+}$ (right) complexes.}
	\label{fig:rdf_PEI10i1_Hg-N}
\end{figure}

We start our analysis of the complexes with the short PEI polymers, PEI-4 and PEI-5, which encircle a Hg$^{2+}$ ion 
within the polymer chains (Figs.~\ref{fig:PEI_model4} and \ref{fig:PEI_model5}). 
The corresponding radial distribution functions, $g(r)$, and running coordination numbers, $n(r)$, are shown in Fig.~\ref{fig:rdf_PEI4_Hg-N}, where 
one can observe that the Hg$^{2+}$ ion associates with $3$ nitrogen atoms of the PEI-4 polymer at the distance of $\sim 2.15$~\AA{} and 
at the same distance to $5$ nitrogen atoms of the PEI-5 polymer. Also, three water molecules enter the first
hydration shell of the ion in the PEI-4+Hg$^{2+}$ complex and one water molecule in the PEI-5+Hg$^{2+}$ complex at the distance of $\sim 2.1$~{\AA} to the ion (Fig. \ref{fig:rdf_PEI4_Hg-Ow}). 
The distributions $g(r)$ and $n(r)$ of nitrogen atoms in PEI polymers (N-N), shown in Fig.~\ref{fig:rdf_PEI4_N-N}, provide insights into the spatial extension of the PEI chain by highlighting characteristic distances between nitrogen atoms, as indicated by the positions of the maxima in $g(r)$. Considering the position of the last maximum in $g(r)$, the largest distance between nitrogen atoms is approximately $4.2$~\AA{} for PEI-5 and $6.4$~\AA{} for PEI-4.

\begin{figure}[!tbh]
	\centering
	\includegraphics[width=0.4\linewidth]{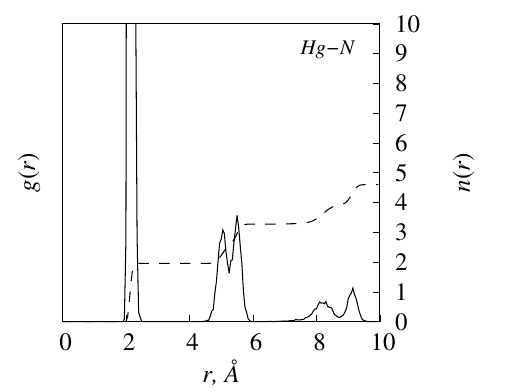}
 	\includegraphics[width=0.4\linewidth]{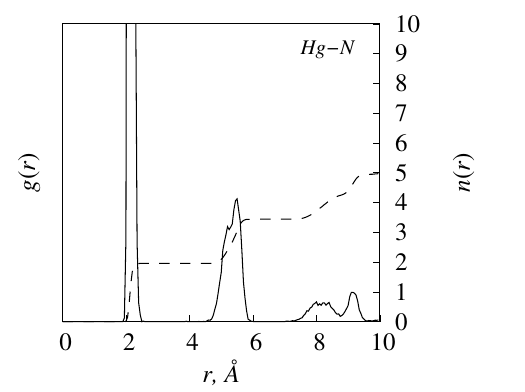}
	\caption{Radial distribution functions $g(r)$ (solid line) and running coordination numbers $n(r)$ (dashed line) Hg-N for PEI-10 + 3~Hg$^{2+}$ (left) and PEI-10 + 4~Hg$^{2+}$ (right) complexes.}
	\label{fig:rdf_PEI10i3_Hg-N}
\end{figure}

\begin{figure}[!tbh]
	\centering
	\includegraphics[width=0.4\linewidth]{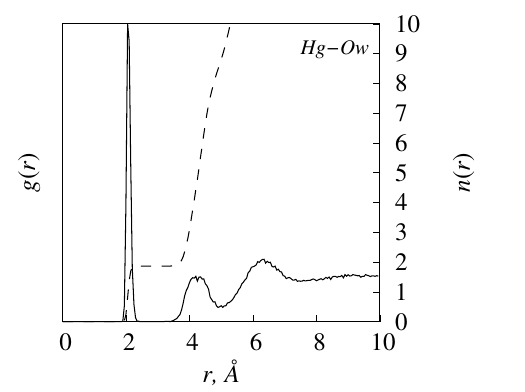}
 	\includegraphics[width=0.4\linewidth]{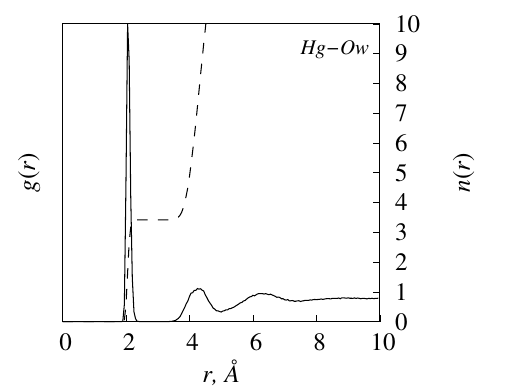}
	\caption{Radial distribution functions $g(r)$ (solid line) and running coordination numbers $n(r)$ (dashed line) Hg-Ow for PEI-10 + 1~Hg$^{2+}$ (left) and PEI-10 + 2~Hg$^{2+}$ (right) complexes.}
	\label{fig:rdf_PEI10i1_Hg-Ow}
\end{figure}

\begin{figure}[!tbh]
	\centering
	\includegraphics[width=0.4\linewidth]{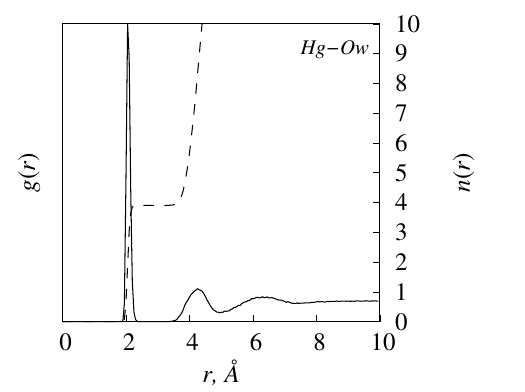}
 	\includegraphics[width=0.4\linewidth]{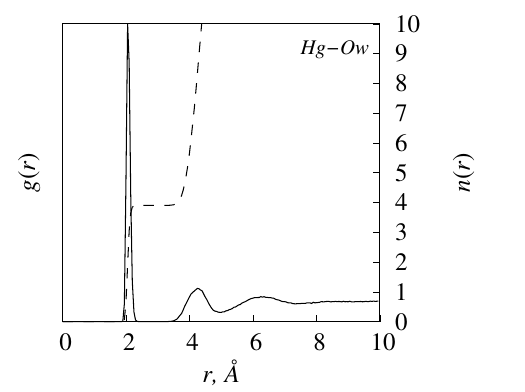}
	\caption{Radial distribution functions $g(r)$ (solid line) and running coordination numbers $n(r)$ (dashed line) Hg-Ow for PEI-10 + 3~Hg$^{2+}$ (left) and PEI-10 + 4~Hg$^{2+}$ (right) complexes.}
	\label{fig:rdf_PEI10i3_Hg-Ow}
\end{figure}

\begin{figure}[!htb]
	\centering
	\includegraphics[width=0.4\linewidth]{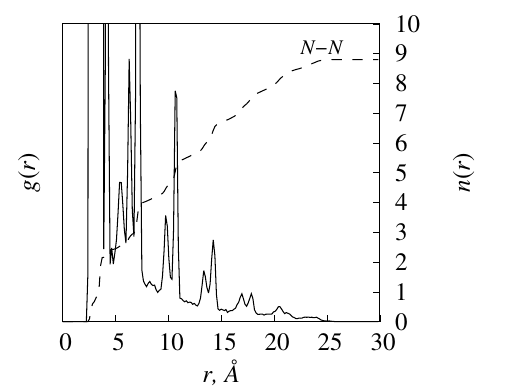}
 	\includegraphics[width=0.4\linewidth]{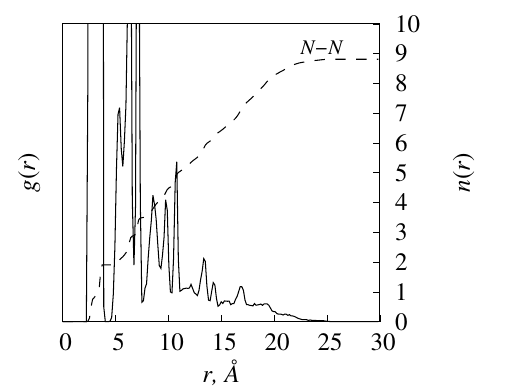}
	\caption{Radial distribution functions $g(r)$ (solid line) and running coordination numbers $n(r)$ (dashed line) N-N for PEI-10 + 1~Hg$^{2+}$ (left) and PEI-10 + 2~Hg$^{2+}$ (right) complexes.}
	\label{fig:rdf_PEI10i1_N-N}
\end{figure}

\begin{figure}[!htb]
	\centering
	\includegraphics[width=0.4\linewidth]{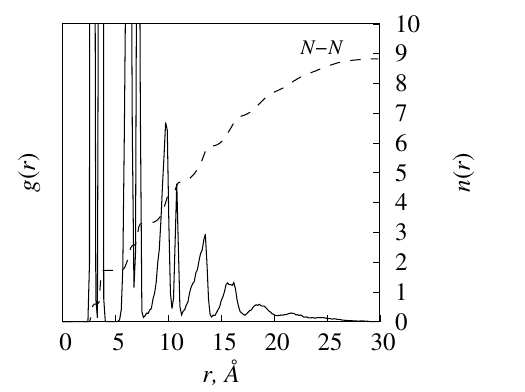}
 	\includegraphics[width=0.4\linewidth]{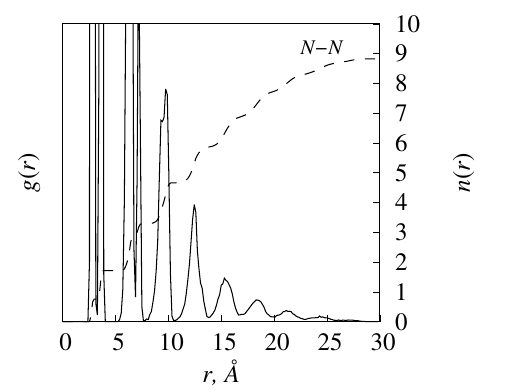}
	\caption{Radial distribution functions $g(r)$ (solid line) and running coordination numbers $n(r)$ (dashed line) N-N for PEI-10 + 3~Hg$^{2+}$ (left) and PEI-10 + 4~Hg$^{2+}$ (right) complexes.}
	\label{fig:rdf_PEI10i3_N-N}
\end{figure}

\begin{figure}[!htb]
	\centering
	\includegraphics[width=0.45\linewidth]{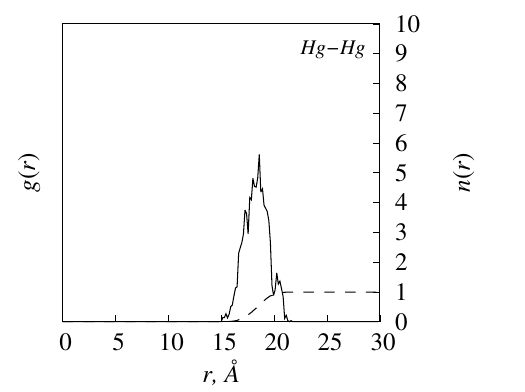}
	\includegraphics[width=0.45\linewidth]{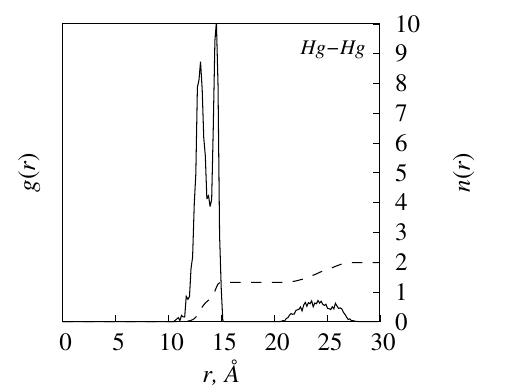}
	\includegraphics[width=0.45\linewidth]{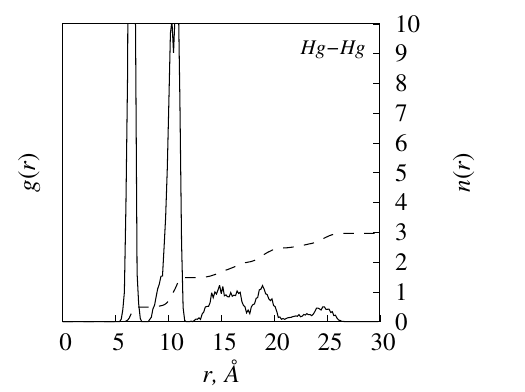}
	\caption{Radial distribution function, $g(r)$ (solid line), and running coordination number, $n(r)$ (dashed line), for Hg--Hg pairs, obtained from MD simulations of the PEI-$10$+$2$~Hg$^{2+}$ (top left), PEI-$10$+$3$~Hg$^{2+}$ (top right),
		and PEI-$10$+$4$~Hg$^{2+}$ (bottom) complexes.
	}
	\label{fig:rdf_Hg-Hg}
\end{figure}

For the longer PEI polymer consisting of $10$ nitrogen atoms (PEI-10) the picture is more complicated, since the complex of PEI-10 + Hg$^{2+}$ involves 
only a part of the polymer chain. One can observe that the complex itself has a shape resembling the case of PEI-5 polymer, but there is also a mobile dangling chain remaining uninvolved (Fig.~\ref{fig:PEI_model10_1}).
The $g(r)$ and $n(r)$ for Hg-N of the PEI-10 + Hg$^{2+}$ complex are shown in Fig.~\ref{fig:rdf_PEI10i1_Hg-N}~(left).  
It is observed that the first $g(r)$ maximum is located in the same position as in the case of PEI-5, but it is slightly smoothed over the larger distance and
only $4$ nitrogen atoms belong to the first coordination shell of mercury allowing one more water molecule to coordinate to the ion (Fig.~\ref{fig:rdf_PEI10i1_Hg-Ow}, left).
The dangling chain of PEI-10 polymer still has the capacity to adsorb other Hg$^{2+}$ ions. 
In Fig.~\ref{fig:rdf_PEI10i1_Hg-N}~(right) we present $g(r)$ and $n(r)$ for the system PEI-10 + 2~Hg$^{2+}$. 
Here, one can see that the first maximum in $g(r)$ of Hg-N is not smoothed anymore. 
It results from a change in conformation of the PEI-10 chain, which is attributed to its reduced mobility in comparison to the free chain due to the Coulomb repulsion 
between two Hg$^{2+}$ ions located at the both ends of the polymer.
The average Hg-N coordination number is $2.5$~(Fig.~\ref{fig:rdf_PEI10i1_Hg-N}), with one ion coordinated to three nitrogen atoms at one end of the chain, and the other coordinated to two nitrogen atoms closer to the opposite end~(Fig.~\ref{fig:PEI_model10_2}).
In systems containing three or four Hg$^{2+}$ ions, no ion is fully encircled by the polymer chain (Figs.~\ref{fig:PEI_model10_3},~\ref{fig:PEI_model10_4}), and the number of neighbouring nitrogen atoms around each Hg$^{2+}$ ion is reduced to two (see Fig.~\ref{fig:rdf_PEI10i3_Hg-N}).
The decrease in the number of nitrogen atoms in the first coordination shell is compensated by water molecules entering the first hydration shell at a distance of approximately $2.1$~\AA{}. For the PEI-10 + 1~Hg$^{2+}$ system, there are $2$ water molecules (Fig.~\ref{fig:rdf_PEI10i1_Hg-Ow}, left; Fig.~\ref{fig:PEI_model10_1}); for PEI-10 + 2~Hg$^{2+}$, there are on average $2.5$ water molecules (Fig.~\ref{fig:rdf_PEI10i1_Hg-Ow}, right; Fig.~\ref{fig:PEI_model10_2}); and for PEI-10 + 3~Hg$^{2+}$ and PEI-10 + 4~Hg$^{2+}$, the number increases to $4$ water molecules in the first hydration shell (Figs.~\ref{fig:rdf_PEI10i3_Hg-Ow}, \ref{fig:PEI_model10_3}, \ref{fig:PEI_model10_4}).
It is worth noting that in the case of Hg$^{2+}$ ion in bulk water, the hydration shell consists of $6$ water molecules, which form an octahedral structure.
The PEI-10 chain is rather elongated in all cases, with the chains chelating $3$ and $4$ Hg$^{2+}$ ions exhibiting the largest maximum distances between nitrogen atoms (see Figs.~\ref{fig:rdf_PEI10i1_N-N}, \ref{fig:rdf_PEI10i3_N-N}). This behaviour can be attributed to electrostatic repulsion between mercury ions and the inherent water solubility of PEI. The minimum distance between Hg$^{2+}$ ions chelated by PEI-10 was approximately $5.55$~\AA{} in the case of $4$ ions, and this minimum distance increased by about $10$~\AA{} as the number of Hg$^{2+}$ ions decreased (Fig.~\ref{fig:rdf_Hg-Hg}).

%%%%%%%%%%%%%%%%%%%%%%%%%%%%%%%%%%%%%%%%%%%%%%%%%%%%%%%%%%%%%%%%%%%%
\section{Validation of obtained structures via density functional theory} \label{sec:5}
%%%%%%%%%%%%%%%%%%%%%%%%%%%%%%%%%%%%%%%%%%%%%%%%%%%%%%%%%%%%%%%%%%%%

We have compared the results observed for the structure of PEI-Hg$^{2+}$ complexes in MD simulations with the structures obtained  
from the density functional theory (DFT) calculations using the Gaussian software (version 16 Rev. C) \cite{frisch2016}. 
Our goal was to determine whether the complexes identified as stable in MD simulations also exhibit stability in density functional theory (DFT) calculations.
For this purpose, we selected several typical configurations of PEI-Hg$^{2+}$ complexes obtained from our MD simulations and used them as input for structure optimization using DFT. The density functionals M06-2X~\cite{zhao2008} were employed in different sets of DFT calculations. The LanL2DZ basis set~\cite{hayInitioEffectiveCore1985} was chosen as suitable for heavy metal ions. In one set of calculations, it was applied to all atoms, while in another set it was used only for Hg$^{2+}$ atoms, with the 6-31+G(d,p) basis set applied to all other atoms.
The presence of water molecules was taken into account implicitly using the solvent-consistent reaction field (SCRF) method. Additionally, geometries of structures that explicitly included the first hydration shells of mercury ions (Fig.~\ref{fig:PEI_models}) were optimized at the M06-2X/LanL2DZ/6-31+G(d,p) level, where the LanL2DZ basis set was applied to mercury atoms and the 6-31+G(d,p) basis set to all other atoms.

\begin{figure}[!htb]
	\centering
	\includegraphics[width=0.14\linewidth]{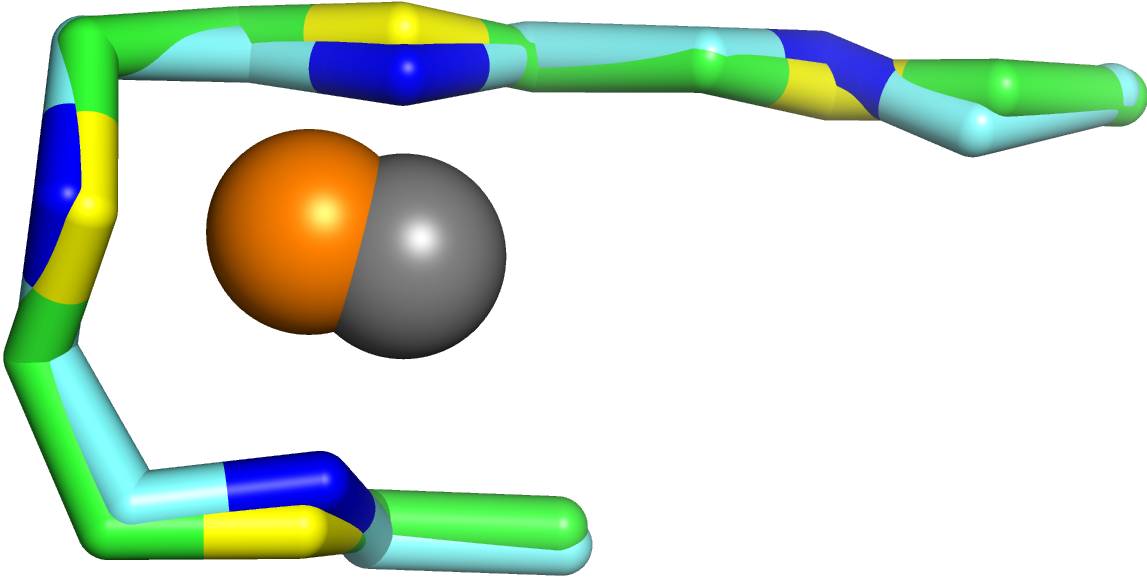}
    \qquad
	\includegraphics[width=0.08\linewidth]{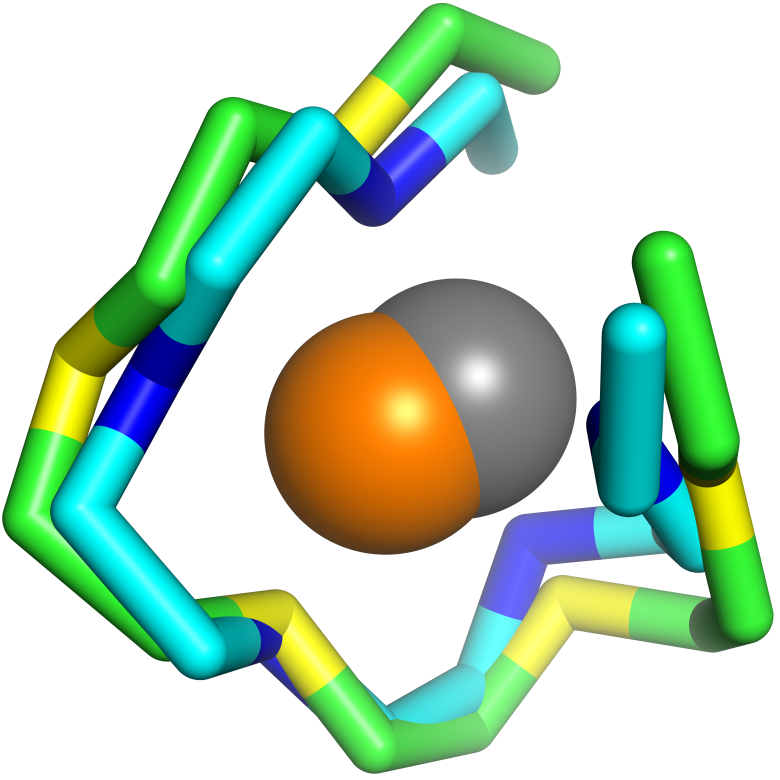}
    \qquad
    \includegraphics[width=0.29\linewidth]{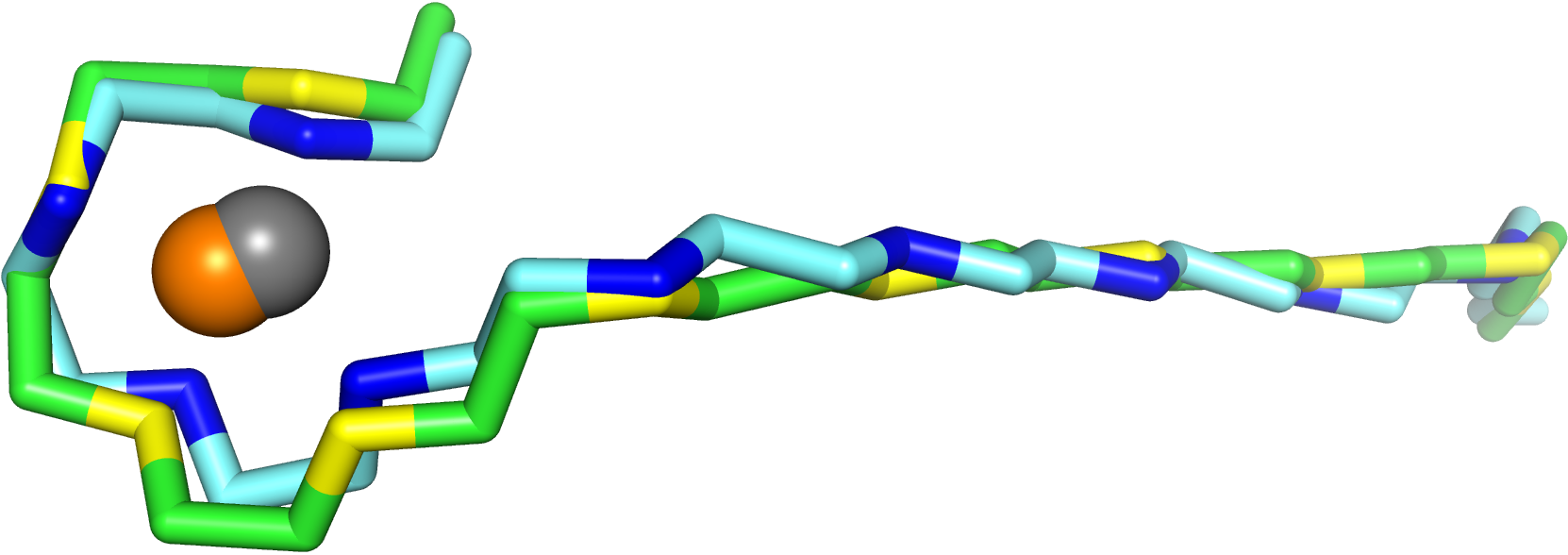}
	\caption{Comparison of DFT and MD conformations of PEI-4, PEI-5 and PEI-10 with 1~Hg$^{2+}$ (MD: Carbon atoms are shown in cyan, nitrogen atoms -- in blue, mercury ion -- in orange, DFT: Carbon atoms are shown in green, nitrogen atoms -- in yellow, mercury ion -- in grey, the pictures were generated using PyMol \cite{pymol}).}
	\label{fig:comparison_md_g}
\end{figure}

\begin{figure}[!htb]
	\centering
	\includegraphics[width=0.46\linewidth]{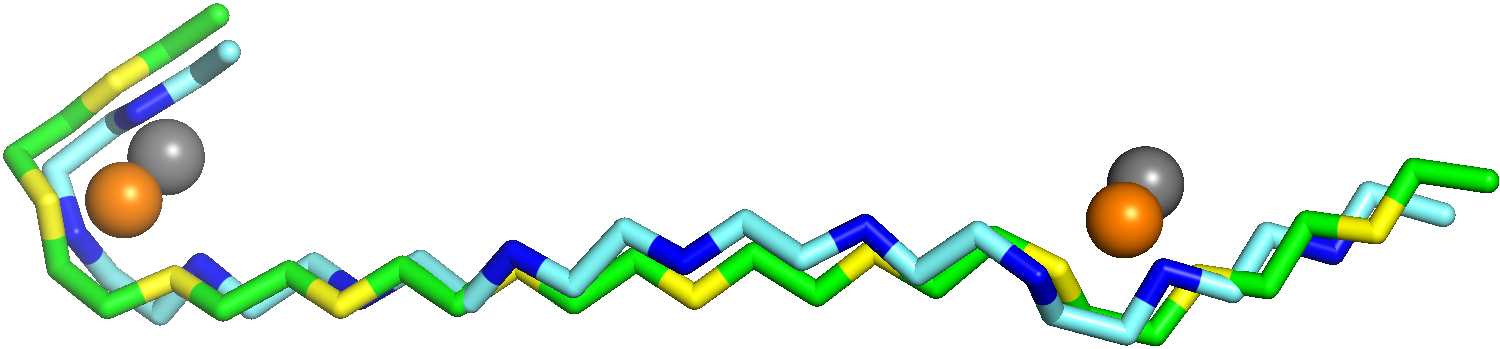}
	\\
	\includegraphics[width=0.5\linewidth]{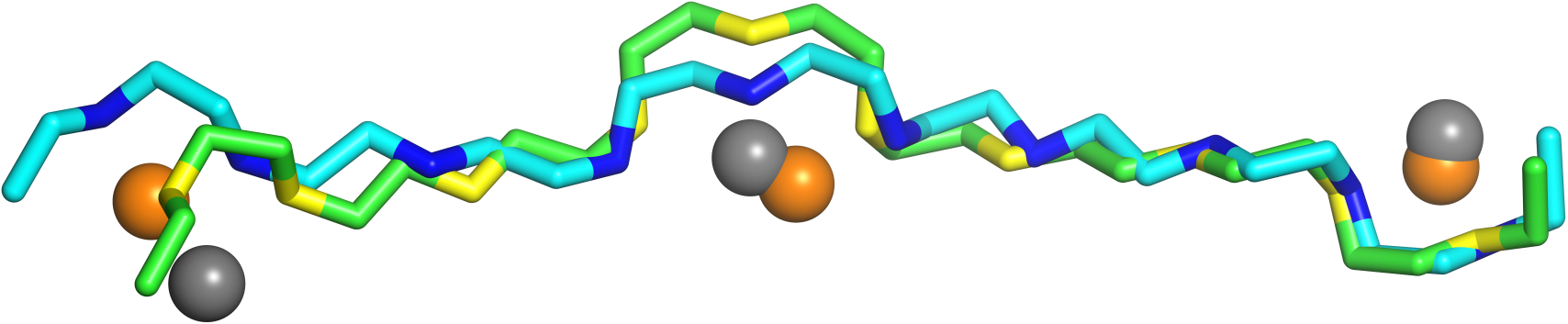}
	\\
	\includegraphics[width=0.48\linewidth]{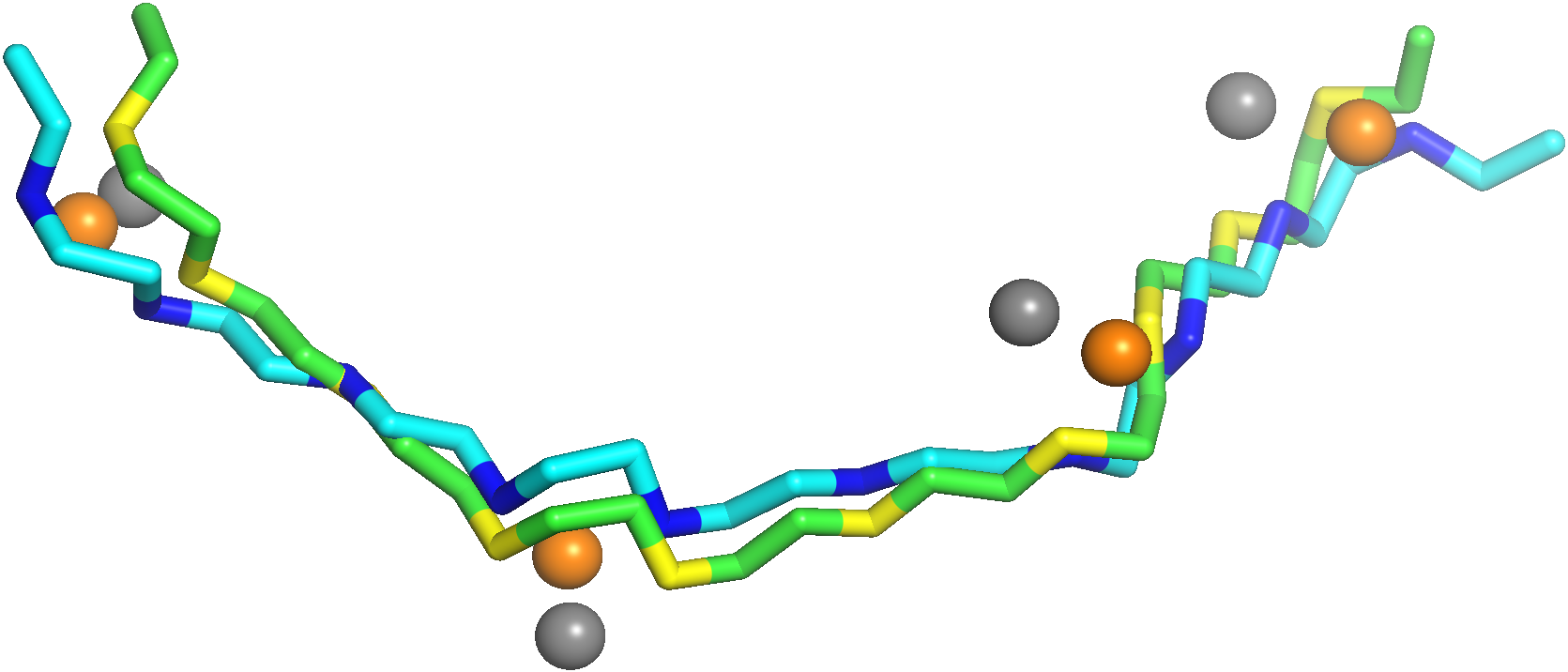}
	\caption{Comparison of DFT and MD conformations of PEI-10 with 2, 3 and 4~Hg$^{2+}$ ions (same colours as in Fig.~\ref{fig:comparison_md_g}).}
	\label{fig:comparisonPei10}
\end{figure}

In Figs.~\ref{fig:comparison_md_g} and \ref{fig:comparisonPei10}, we present a comparison between the structures of complexes obtained from MD simulations and those optimized using DFT. In each panel of these figures, the MD and DFT structures are shown aligned for best fit. As can be seen, both approaches yield structures that are in good qualitative agreement. In particular, for complexes containing a single Hg$^{2+}$ ion coordinated with PEI-4, PEI-5, and PEI-10 (Fig.~\ref{fig:comparison_md_g}), the DFT-optimized structures closely reproduce those obtained from MD simulations. The overall shapes of the complexes are preserved, and the number of nitrogen atoms in the first coordination shell of the Hg$^{2+}$ ion is also equivalent.

There is considerable interest in the case where two or more mercury ions are complexed with the PEI-10 molecule. Similarly to our MD simulations, we have examined complexes containing PEI-10 with $2$, $3$, and $4$~Hg$^{2+}$ ions. As shown in Fig.~\ref{fig:comparisonPei10}, these complexes also proved to be stable in our DFT calculations, with their optimized structures closely resembling those obtained from MD simulations. Minor discrepancies are primarily due to the thermal motion of polymer segments in the MD simulations. Nevertheless, the overall conformations remain consistent between both approaches. It is worth noting that a further increase in the number of Hg$^{2+}$ ions involved in the complexation with the PEI-10 chain does not result in a stable complex, indicating that the chelating capacity of the PEI-10 polymer is limited to four ions.

Despite a close similarity of PEI conformations in the PEI+Hg$^{2+}$ complex, there is still visible difference of the Hg$^{2+}$ ion positions obtained by the MD and DFT methods. A small shift in respect to the polymer, which is seen almost in all pictures of Fig.~\ref{fig:comparison_md_g} and \ref{fig:comparisonPei10} indicates the different distances between Hg$^{2+}$ ion and nitrogen atoms as provided by these methods.
Thus, we have also checked the distances between Hg$^{2+}$ ions and nitrogen atoms of the PEI molecule in the complexes calculated from DFT 
and compared them with those obtained from MD based on the radial distribution functions of Hg-N, which were discussed above. 
It is observed that the average Hg-N distances obtained from MD simulations are smaller than those calculated from the DFT-optimized structures (Table~\ref{t:gau_dist}), which are approximately $2.5$~\AA{} in most cases.

\begin{table}[!htb]
	\begin{center}
		\caption{\label{t:gau_dist} The distances between Hg$^{2+}$~ions and N~atoms of PEI involved in complex~(\AA{}). The complex structures were optimized at the M06-2X/LanL2DZ/6-31+G(d,p) level.}
		\begin{tabular}{p{3.3em} p{0.2em} p{2.5em} p{2em} p{2em} p{2em} p{2em} p{2em} p{2em} p{2em} p{2em} p{2em} p{2em}}
\hline\hline
\multicolumn{3}{c}{Complex} & Hg$^{2+}$ & N1   & N2   & N3   & N4   & N5   & O$_1$ & O$_2$ & O$_3$ & O$_4$ \\ \hline\hline
PEI-4  & + & 1~Hg$^{2+}$    & 1         & 2.49 & 2.51 & 2.57 &  &      & 2.55    & 2.66   &   2.7     &        \\ \hline  
PEI-5  & + & 1~Hg$^{2+}$    & 1         & 2.48 & 2.51 & 2.53 & 2.56 & 2.65 & 3.19   &        &        &        \\ \hline
PEI-10 & + & 1~Hg$^{2+}$    & 1       & 2.47 &  2.49& 2.55 & 2.57       &      &   2.6 &  2.73  &     &        \\ \hline
PEI-10 & + & 2~Hg$^{2+}$    & 1         & 2.49 & 2.49 & 2.57 &      &      & 2.59   & 2.6    & 2.65   &        \\ \hline
&   &                & 2         & 2.48 & 2.51 &      &      &    & 2.53 & 2.54  & 2.55    &  2.56           \\ \hline
PEI-10 & + & 3~Hg$^{2+}$    & 1         & 2.5  & 2.54 &      &      &      & 2.53   & 2.54   & 2.56   & 2.56   \\ \hline
&   &                & 2         & 2.49 & 2.57 &      &      &      & 2.5    & 2.55   & 2.56   & 2.56   \\ \hline
&   &                & 3         & 2.52 & 2.53 &      &      &      & 2.51   & 2.54   & 2.56   & 2.59   \\ \hline
PEI-10 & + & 4~Hg$^{2+}$    & 1         & 2.51 & 2.55 &      &      &      & 2.48   & 2.49   & 2.5    & 2.63   \\ \hline
&   &                & 2         & 2.5  & 2.52 &      &      &      & 2.48   & 2.54   & 2.54   & 2.62   \\ \hline
&   &                & 3         & 2.49 & 2.54 &      &      &      & 2.51   & 2.55   & 2.55   & 2.72   \\ \hline
&   &                & 4         & 2.49 & 2.52 &      &      &      & 2.55   & 2.59   & 2.59   & 2.61   \\ \hline\hline
		\end{tabular}
	\end{center}
\end{table}

\section{Adsorption energy} \label{sec:6}

To assess whether the complex formation is thermodynamically favourable, adsorption energies and Gibbs free energies of complexation were calculated following MD simulations and DFT geometry optimizations, respectively. The adsorption energy represents the energy change associated with the binding of a metal ion to a chelating molecule. It was calculated as the difference between the total energy of the complex and the sum of the energies of the isolated chelating agent and the unbound metal ion.

The adsorption energies were calculated for the complexes according to the expression $E_{\text{cmpx}} = E(\text{PEI}+\text{Hg}^{2+}) - E(\text{PEI}) - E(\text{Hg}^{2+})$, where $E(\text{PEI}+\text{Hg}^{2+})$, $E(\text{PEI})$, and $E(\text{Hg}^{2+})$ denote the potential energy of the PEI-Hg$^{2+}$ complex, the potential energy of PEI in water, and the potential energy of unbound (dechelated) Hg$^{2+}$ in water, respectively.
To obtain $E(\text{PEI})$ and $E(\text{Hg}^{2+})$, additional MD simulations were performed for free PEI polymers (PEI-$4$, PEI-$5$, and PEI-$10$) and the unbound Hg$^{2+}$ ion in an aqueous environment, following the MD protocol described earlier.
The results indicate favourable complexation (Table~\ref{table:EADPEI}), with the strongest mercury ion binding observed for the PEI-$5$ polymer, followed by PEI-$10$ and PEI-$4$ chelating a single ion, and then by PEI-$10$ chelating $2$, $3$, and $4$ ions.

\begin{table*}[!htb]
	\centering
	\caption{\emph{G$_{cmpx}$} (kJ mol$^{-1}$) per Hg $^{2+}$ ion for the PEI-mercury complexes.
		{Water is taken into account implicitly.}} %complexation Gibbs free energies
	\label{table:EADPEI}
	\begin{tabular}[t]{p{47pt}p{0.5pt}lrp{2pt}ccccc}
		\hline\hline
		\multicolumn{3}{c}{System} & \multicolumn{3}{c}{MD} & \makecell{Binding\\enegry, MD} & \makecell{M06-2X/\\LanL2DZ} & \makecell{M06-2X/\\LanL2DZ/\\6-31+g(d,p)} \\ \hline\hline
		PEI-4  & + & 1 Hg$^{2+}$   & -58 &  &    &       -395$\pm$15        &            {-131}             &                   {-47}                                     \\
		PEI-5  & + & 1 Hg$^{2+}$   &  -112 &  &    &       -670$\pm$21        &          -174             &                   -70                 \\
		PEI-10 & + & 1 Hg$^{2+}$   & -91 &  &       &       -541$\pm$58        &         {-169}            &                  {-61}                     \\
		PEI-10 & + & 2 Hg$^{2+}$  & -51  &  &    &       -333$\pm$12        &           {-97}            &                  {-38}                             \\
		PEI-10 & + & 3 Hg$^{2+}$   & -40  &  &    &       -258$\pm$8       &            -82             &                   -30                      \\
		PEI-10 & + & 4 Hg$^{2+}$   & -29 &  &    &       -256$\pm$8        &            -65             &                   -21                    \\
		\hline\hline
	\end{tabular}
\end{table*}

\begin{figure}[!tbh]
	\centering
	\includegraphics[width=0.7\linewidth]{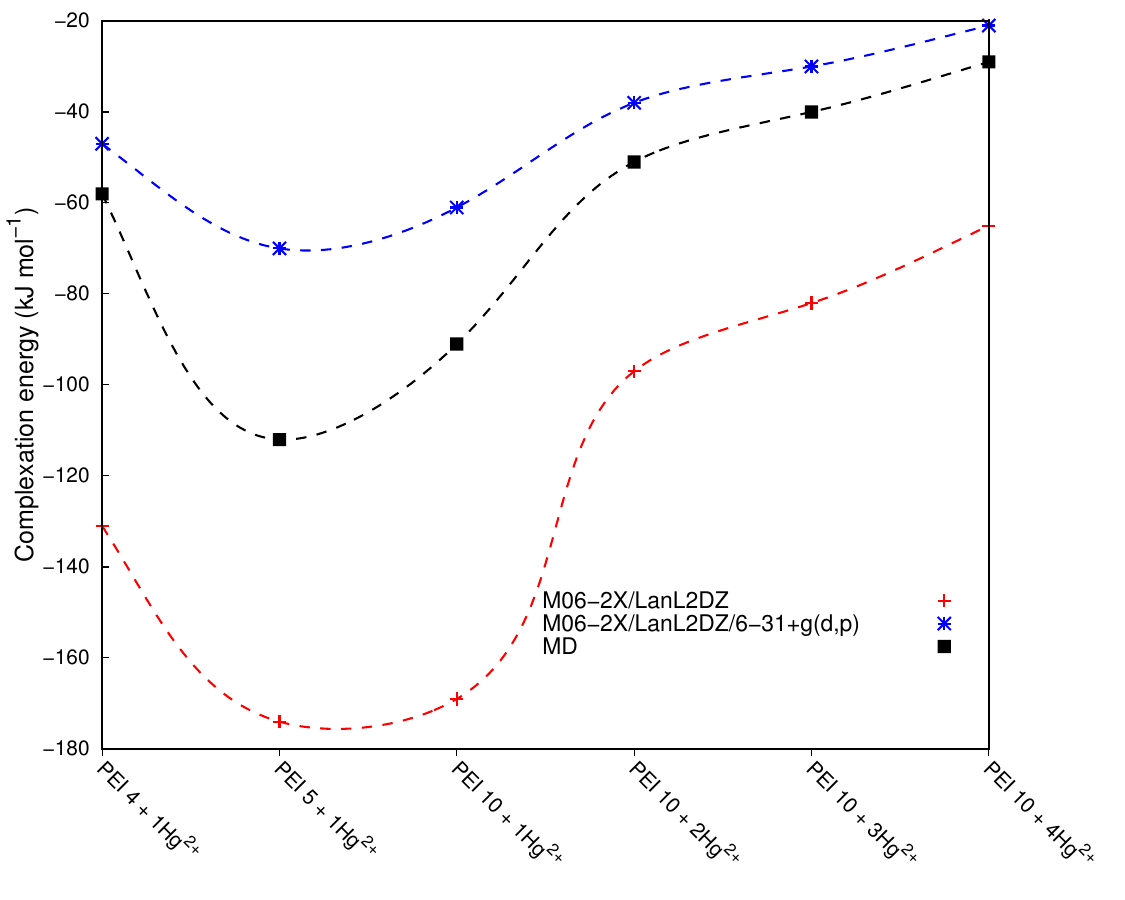}
	\caption{\label{fig:pic_e} Complexation energies per Hg $^{2+}$ ion (see Table~\ref{table:EADPEI}). The aqueous environment is taken into account implicitly. The dashed lines help to guide the eye.}
\end{figure}

For all systems and levels of theory used, the Gibbs free energies of complexation were calculated using the expression  
$G_{\text{cmpx}} = G(\text{PEI} + \text{Hg}^{2+}) - G(\text{PEI}) - G(\text{Hg}^{2+})$,  
where $G$ denotes the sum of electronic and thermal free energies at $T = 298.15$~K.  
To compute $G(\text{PEI})$, the geometry of the PEI molecule was first optimized using the corresponding level of theory. The initial PEI structure for optimization was taken from an additional MD simulation of the free polymer.  
Negative values of $G_{\text{cmpx}}$ indicate thermodynamically favourable complexation for all studied cases (Table~\ref{table:EADPEI}). The results show stronger binding of a single Hg$^{2+}$ ion in the case of PEI-5 compared to PEI-4 at all computational levels (Fig.~\ref{fig:pic_e}), and compared to all other systems when using levels based on the M06-2X functional. This can be attributed to the largest Hg-N coordination number observed for the Hg$^{2+}$-PEI-5 complex.  
As the number of mercury ions increases, the complexation energy becomes less negative, indicating progressively weaker binding.

The results show that PEI-5 forms a more stable complex with a single Hg$^{2+}$ ion than PEI-4 across all computational levels (Fig.~\ref{fig:pic_e}), and demonstrates the strongest binding among all studied systems when using levels of theory based on the M06-2X functional. This enhanced stability can be attributed to the highest Hg-N coordination number observed in the Hg$^{2+}$-PEI-5 complex. As the number of mercury ions in the system increases, the complexation energy becomes progressively less negative, indicating a reduction in binding strength.

%%%%%%%%%%%%%%%%%%%%%%
\section*{Conclusions}
%%%%%%%%%%%%%%%%%%%%%%

We have studied the chelation of mercury ions Hg$^{2+}$ by linear PEI polymers in an aqueous environment. 
Atomistic computer simulations are performed to offer a microscopic description of the complexation between an isolated PEI chain and Hg$^{2+}$. 
For this purpose, the MD method is applied in combination with the OPLS/AA-based force field for different lengths, $k$, of PEI polymers, 
where $k$ denotes the number of nitrogen atoms in a considered polymer chain. 
The study focused on two aspects of the complexation: (i)~stability of the pre-initiated complexes formed by a single PEI-$k$ polymer and Hg$^{2+}$ ions 
depending on the polymer length $k$ and the number of ions involved, and (ii)~spatial structure of the stabilized complexes. 
The latter was determined via evaluation of the radial distribution functions for the Hg-N and N-N atomic pairs.
For the short chains, PEI-4 or PEI-5, involving one Hg$^{2+}$ ion, the polymer is characterized by a highly bend conformation with the prevailance of non-\textit{trans}-dihedral angles, whereas an ion is highly dehydrated. The complexation in this case is caused by the association occurring between nitrogen atoms and a Hg$^{2+}$ ion. For the longer chain, PEI-10, stable complexes are formed with up to four ions per chain. In these cases, the ions are moderately hydrated, and they are separated by stretched polymer fragments with predominance of \textit{trans} dihedral angles. Therefore, the complexation here is affected by Coulombic repulsion between the Hg$^{2+}$ ions. The complexes obtained from the MD simulations were studied using the DFT calculations. 
The latter confirmed the stability of the observed conformations for the PEI chains and spatial arrangements of the mercury ions in these complexes.

Future extensions of this work may involve more refinement of the atomistic PEI model, e.g. using alternative force fields; its coarse-graining aimed at covering larger length scales; as well as the application of suggested approach to numerous cases of branched PEI architectures.

%%%%%%%%%%%%%%%%%%%%%%
\section*{Acknowledgements}
%%%%%%%%%%%%%%%%%%%%%%

H.B. expresses gratitude to the Finnish National Agency for Education for financial support within the EDUFI Fellowship for doctoral students from Ukraine (ref. no. OPH-4602-2022) and the Team Finland Knowledge programme (ref. no. 12/221/2023),
CSC (Finnish IT Center for Science) is acknowledged for providing computing resources.
T.P. and J.I. gratefully acknowledges the financial support from the National Research Foundation of Ukraine (grant no. 2023.05/0019)
and ICCS (Interdisciplinary Center for Computing Simulations, Lviv) for additional computing resources.

%%%%%%%%%%%%%%%%%%%%%%
\section{Data Availability}
%%%%%%%%%%%%%%%%%%%%%%
Data will be made available on request.

%\renewcommand{\refname}{References}
%\def\bibsection{\section*{\refname}}
%\bibliographystyle{elsarticle-num}
%\bibliography{PEI.bib}

\end{document}